\documentclass[acmsmall,screen]{acmart}

\usepackage{amsmath,amsfonts}
\usepackage{algorithmic}
\usepackage{graphicx}
\usepackage{textcomp}
\usepackage{xcolor}
\usepackage{comment}
\usepackage{balance}
\usepackage{xspace}
\usepackage[ruled, lined, linesnumbered, commentsnumbered, longend]{algorithm2e}
\usepackage{multirow,multicol}
\usepackage{colortbl}
\usepackage{tabulary}
\usepackage[flushleft]{threeparttable}
\usepackage{caption}
\usepackage{subfigure}
\usepackage{tcolorbox}
\usepackage{mathtools}
\usepackage{url}
\usepackage{circledsteps}
\usepackage{tikz}
\usepackage{units}
\usepackage{diagbox}
\usepackage{listings}
\usepackage{enumitem}
\definecolor{backcolour}{rgb}{0.95,0.95,0.92}

\SetCommentSty{mycommfont}

\newcommand{\rqone}{\textit{RQ1: What is the quality of API-oriented code generation by LLMs?}\xspace}

\newcommand{\rqtwo}{\textit{RQ2: What types of errors occur in API-oriented code generation?}\xspace}

\newcommand{\rqthree}{\textit{RQ3: What factors are associated with the quality of API-oriented code generation by LLMs?}\xspace} 

\newcommand{\rqfour}{\textit{RQ4: Does RAG help mitigate the errors in API-oriented code generation?}\xspace}

\newcommand{\rqboxc}[1]{\begin{tcolorbox}[left=1pt,right=1pt,top=0pt,bottom=0pt,colback=gray!5,colframe=gray!40!black,before skip=5pt,after skip=0pt]#1\end{tcolorbox}}

\usepackage{tabularx}
\usepackage{listings}
\usepackage{color}

\definecolor{dkgreen}{rgb}{0,0.6,0}
\definecolor{gray}{rgb}{0.5,0.5,0.5}
\definecolor{mauve}{rgb}{0.58,0,0.82}

\usepackage{graphicx}
\graphicspath{ {./figures/} }

\usepackage{textgreek}
\newcommand{\phead}[1]{\noindent {\bf #1}}

\usepackage{float}

\newcommand{\NoneExistP}{IncorrectAPI\xspace} 

\newcommand{\NoAPIUsedP}{NoAPIInvoked\xspace} 

\newcommand{\UnCompilableP}{Uncompilable\xspace} 
\newcommand{\NonRunnableP}{Unexecutable\xspace} 

\newcommand{\total}{TotalError\xspace}

\newcommand{\ourtool}{\textsc{AutoAPIEval}\xspace}

\AtBeginDocument{%
  \providecommand\BibTeX{{%
    \normalfont B\kern-0.5em{\scshape i\kern-0.25em b}\kern-0.8em\TeX}}}

\begin{document}


\title{A Comprehensive Framework for Evaluating API-oriented Code Generation in Large Language Models}

\author{Yixi Wu}
\affiliation{%
  \institution{University of Manitoba}
  \city{Winnipeg}
  \country{Canada}}
\email{wuy8@myumanitoba.ca}

\author{Pengfei He}
\affiliation{%
  \institution{University of Manitoba}
  \city{Winnipeg}
  \country{Canada}}
\email{hep2@myumanitoba.ca}

\author{Zehao Wang}
\affiliation{%
  \institution{Concordia University}
  \city{Montreal}
  \country{Canada}}
\email{w_zeha@encs.concordia.ca}

\author{Shaowei Wang}
\affiliation{%
  \institution{University of Manitoba}
  \city{Winnipeg}
  \country{Canada}}
\email{shaowei.wang@umanitoba.ca}

\author{Yuan Tian}
\affiliation{%
  \institution{Queen's University}
  \city{Kingston}
  \country{Canada}}
\email{y.tian@queensu.ca}

\author{Tse-Hsun (Peter) Chen}
\affiliation{%
  \institution{Concordia University}
  \city{Montreal}
  \country{Canada}}
\email{peterc@encs.concordia.ca}

\begin{abstract}
Large language models (LLMs) like GitHub Copilot and ChatGPT have emerged as powerful tools for code generation, significantly enhancing productivity and accelerating software development. However, existing benchmarks primarily focus on general code generation without considering API-oriented code generation, i.e., generating code that invokes APIs from specific libraries. Given the growing demand for API-oriented code generation, there is a pressing need for a systematic and automated approach to evaluate LLM on API-oriented code generation. To address this gap, we propose \ourtool, a lightweight and automated framework designed to evaluate the capabilities of LLMs in API-oriented code generation. Our framework works with any library that provides API documentation and focuses on two unit tasks: API recommendation and code example generation, along with four metrics to evaluate the generated APIs and code examples, such as the proportion of incorrect API recommendations for Task 1, and the proportion of code examples where no specific API is invoked and uncompilable/unexecutable code examples for Task 2.
In addition, we conducted a case study on three LLMs (ChatGPT, MagiCoder, and DeepSeek Coder) and Java Runtime Environment 8 to demonstrate the framework's effectiveness. Our findings reveal substantial variability in LLM performance across tasks, with ChatGPT adhering better to instructions, while sharing similar effectiveness in code example generation with its counterparts (i.e., MagiCoder and DeekSeek Coder). We also identify key factors associated with code quality, such as API popularity and model confidence, and build classifiers that achieve high accuracy in detecting incorrect API recommendations and erroneous code examples. Retrieval-augmented generation enhances the quality of code generated by LLMs, though its effectiveness varies across different LLMs. 
\end{abstract}

%


\maketitle

\section{Introduction}\label{sec:intro}

With the technological advancements in AI, various large language models (LLMs) have been developed for code-related tasks~\cite{codexglue,codex,wang2021codet5,guo2022unixcoder,luo2023wizardcoder,wei2023magicoder,li2023starcoder}, such as GitHub Copilot~\cite{GithubCopilot} and ChatGPT~\cite{ChatGPT}. Those LLMs have significantly propelled the field of code generation. When used correctly, those LLMs can significantly enhance productivity and accelerate software development through various tasks~\cite{peng2023impact,paredes2023chatgpt,dakhel2023github}. To address the growing demand for LLM-based code generation, prior studies have devoted substantial effort to evaluating and understanding the effectiveness of LLMs in generating code~\cite{dou2024s,liu2024exploring,spracklen2024we,human-eval,du2023classeval}. For instance, to assess LLM's ability of code generation, researchers proposed various code generation benchmarks (e.g., HumanEval~\cite{human-eval} and ClassEval~\cite{du2023classeval}). These benchmarks rely on a set of pre-defined tests to evaluate the correctness of the generated code.




According to the latest survey on developers' perspectives regarding code generation tools~\cite{ciniselli2023source}, developers frequently expect these tools to incorporate more contextual knowledge, particularly by leveraging specific libraries. They emphasize the importance of \textit{API-oriented code generation}, where tools should be capable of generating code that invokes APIs from specific libraries. However, the majority of the existing studies~\cite{khoury2023secure, dou2024s,liu2024exploring,spracklen2024we,human-eval,du2023classeval} focus on the general code generation based on specific functional requirements described in natural language, without specifying APIs. Little work has been done to assess the quality of API-oriented code generated by LLMs. 
While some research has explored API-oriented code generation for specific libraries~\cite{codegen4libs, zan2023private}, these studies typically derive their tasks from Stack Overflow or general code generation benchmarks, which involve APIs from the target libraries (e.g., Pandas). They then manually create test cases to verify if the generated code meets the required functionality. However, a key limitation of these approaches is that the coverage of tested APIs is limited and cannot be scaled to a broader range of libraries due to the manual effort required for test creation and the difficulty of collecting relevant tasks for the target libraries. Hence, a systematic and automated approach is needed to evaluate API-oriented code generation by LLMs for a broad range of libraries. Enabling the assessment of API-oriented code generation can help practitioners evaluate and analyze the code generation capabilities of LLMs for specific libraries, thereby offering new insights and solutions to enhance their performance.

To bridge this gap, we propose a lightweight framework \ourtool, which enables automatic and systematic evaluation of LLMs on API-oriented code generation. \ourtool works with any library that has API documentation. We consider four requirements when designing \ourtool: First, it should be fully automated to enable scalability, unlike existing benchmarks that require manually crafted test cases~\cite{human-eval,du2023classeval}. Second, it should be applicable to a wide range of libraries, leveraging easily accessible datasets such as API documentation. Third, the framework's tasks must be simple enough for LLMs to understand without complex prompt engineering, ensuring consistent benchmarking~\cite{hajipour2023codelmsecbenchmark}. Lastly, the framework should mimic how developers interact with LLMs~\cite{hao2024empiricalstudydevelopersshared}, helping them identify available APIs and providing examples of working code.

To fulfill the requirements, we design two unit tasks in \ourtool to benchmark LLMs' ability in API-oriented code generation, using only API documentation as input. We refer to them as ``unit tasks'' because, similar to unit tests, they are performed independently on each class within a package, allowing for focused evaluation of the LLM's performance at the class level. The tasks are: 1) \textbf{API recommendation}, and 2) \textbf{code example generation}, evaluated using four specific metrics. Specifically, in Task 1, given a library, we iteratively query the LLM to recommend a list of APIs for each class (i.e., unit) in the library. In Task 2, we query the LLM to generate code examples for a given API. We propose four evaluation metrics to evaluate the API and code examples generated by LLMs, such as the proportion of incorrect APIs for Task 1, the proportion of code examples where no specific API is invoked, and uncompilable/unexecutable code examples for Task 2. 
Our framework enables further analysis to 1) understand errors that occur in the API recommendation and code generation tasks; 2) investigate factors that are associated with the quality of APIs recommended and code examples generated by LLMs; and 3) investigate potential solutions to mitigate errors and improve the quality APIs and code examples generated by LLMs. For simplicity, we refer to the API recommendation in Task 1 and code example generation in Task 2 as \textit{API-oriented code generation} below. Similarly, we refer to the APIs and code examples generated by LLMs as \textit{API-oriented code}.

To illustrate how our framework can be used for further analysis, we conducted a case study on Java Runtime Environment 8 (JRE 8). We collected the API documentation of JRE 8, which comprises 217 packages and 2,397 classes. We selected three popular LLMs for evaluation, including one state-of-the-art closed-source LLM, ChatGPT~\cite{ChatGPT}, and two open-source LLMs that are trained for code-related tasks, MagiCoder~\cite{magicoder} and DeepSeek Coder~\cite{deepseek-coder}.

    
We made the following observations through the case study:
\vspace{-0.1cm}
\begin{itemize}
    \item  Different LLMs vary in performance across different tasks. Compared with MagiCoder and DeepSeek Coder, ChatGPT tends to follow the instructions better and have a lower rate of hallucinations, while producing a similar proportion of uncompilable/unexecutable code examples.
    \item For API recommendation, 58.1\% to 84.1\%  of recommended APIs do not exist in the specified library. In code example generation, 39.4\% to 54.4\% of examples contain errors, with 5.4\% to 20.7\% missing the specified API and the rest failing to compile or execute. To further understand the errors, we created a taxonomy of the errors that occurred in the API-oriented code generation by LLMs. For the API recommendation task, most errors stem from Factual Fabrication Hallucinations, followed by Instruction Inconsistencies, and Factual Inconsistencies. In code example generation, errors are primarily due to Runtime Errors (e.g., Initialization Error), followed by Hallucinations and Compilation Errors.
    \item Factors such as API popularity and model confidence are strongly associated with API-oriented code quality. Using our proposed factors, we built highly accurate classifiers to detect incorrect API recommendations and erroneous code examples (e.g., F1 scores of 0.96 and 0.8 for Task 1 and 2 on MagiCoder).
0.79 for Task 1 and Task 2 using MagiCoder.
    \item Overall, RAG enhances the quality of code generated by LLMs, though its effectiveness varies across different models. For Task 1, even when provided with a list of correct APIs, LLMs still exhibit an error rate of at least 27.9\% in their recommendations.
\end{itemize}

We summarize our contributions below:

\begin{itemize}
    \item We proposed an automated and systematic framework to enable the evaluation of LLMs on API-oriented code generation and various further analyses. 
    \item We conducted a case study on three LLMs and 217 packages from JRE 8 using our framework, uncovering key insights that suggest valuable insights for future research. 
    \item We release our replication package~\cite{AutoAPIEval} to facilitate future research. 
    
\end{itemize}

\section{\ourtool: Framework for Evaluating API-oriented Code Generation}\label{sec:framework}

We propose a lightweight framework named \ourtool. This framework enables the automatic and systematic evaluation of LLMs for API-oriented code generation, which is applicable to any specific library and requires only API documentation as input. 

We consider four key requirements when designing the framework.
First, \textbf{Fully Automated:} The evaluation must be conducted automatically. Without full automation, scaling to new libraries is impractical. Existing benchmarks, such as HumanEval~\cite{human-eval} and ClassEval~\cite{du2023classeval}, rely on predefined test cases to assess the correctness of generated code. While these benchmarks are meticulously crafted, creating test cases requires substantial manual effort, making it challenging to extend them to new libraries.
Second, \textbf{Easily Accessible Dataset:} To ensure our framework's applicability across a wide range of libraries, the datasets required for evaluation must be readily accessible.
Third, \textbf{ Easy to Implement:} The tasks in our framework should be straightforward so that LLMs can easily understand the intent of the prompts. Complex tasks often require intricate prompt engineering, which does not support a standardized approach for benchmarking LLMs~\cite{hajipour2023codelmsecbenchmark}.
Lastly, \textbf{Mimic Practical Use:} The tasks should emulate how developers interact with LLMs for API-oriented code generation. Typically, developers start by querying LLMs about the APIs available in a library and requesting usage examples before diving deeper into specific tasks~\cite{hao2024empiricalstudydevelopersshared}.

\begin{figure}
    \centering
    \includegraphics[width=1\linewidth]{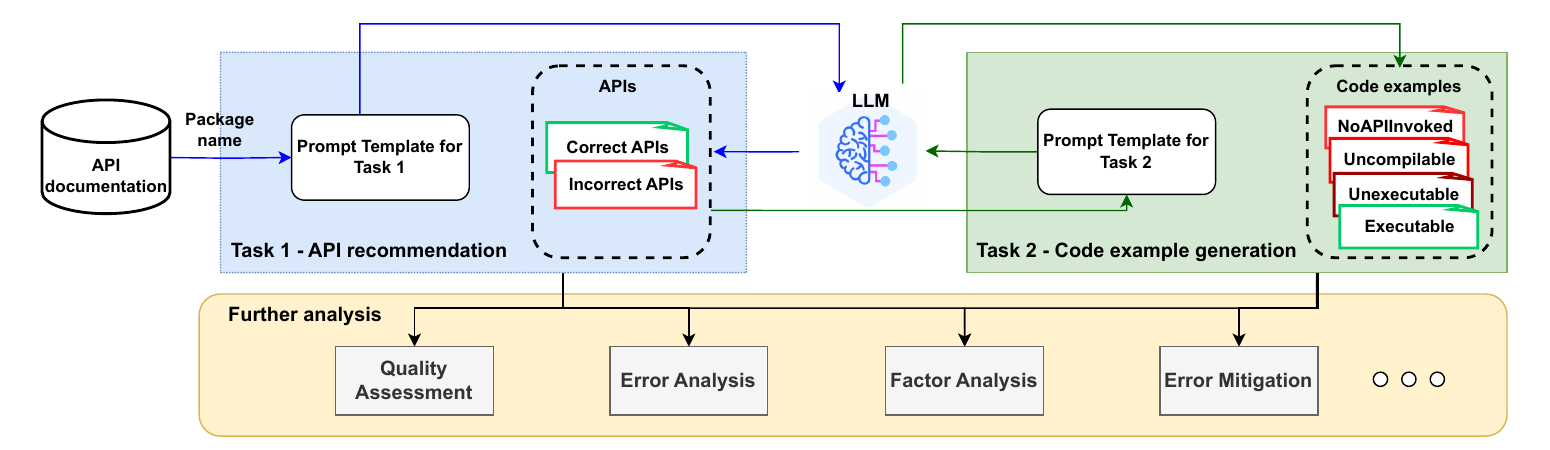}
    \caption{Framework of \ourtool.}
    \label{fig:framework}
    \vspace{-0.3cm}
\end{figure}

Figure~\ref{fig:framework} provides an overview of the proposed framework for assessing the API-oriented code generated by LLMs. To fulfill the four requirements, we design two tasks in \ourtool for evaluating LLMs' ability in API-oriented code generation: \textbf{API recommendation} and \textbf{code example generation}. We also propose four evaluation metrics, which can be used to evaluate the quality of the recommended APIs for Task 1 and the generated code examples for Task 2, such as the proportion of correctly recommended APIs and the proportion of code examples that are uncompilable or unexecutable (see more details in Section~\ref{sec:evaluationMetrics}).

In addition, our framework can facilitate further analysis, including 1) identifying errors in the generated code, 2) determining factors and indicators potentially related to code quality, and 3) evaluating solutions to mitigate errors in the generated code. We demonstrate how to apply our \ourtool to conduct such analyses in Section~\ref{sec:experimentalsetting} via a case study. 


\subsection{Tasks}\label{sec:task}

Given a library, in Task 1, the LLM is iteratively prompted to recommend methods for each class in the library. In Task 2, the LLM generates code examples for a specified method in the given library. Note that Task 2 is only applied to methods that were successfully recommended by the LLM in Task 1, as we assume that the LLM has adequate knowledge of these APIs. We elaborate on those two tasks in the following sub-sections.

\subsubsection{Task 1: API Recommendation}\label{sec:task1}
For Task 1, we test LLM's knowledge of the available APIs in a specific library. Therefore, we prompt the LLM to recommend APIs (methods) in a given class in the specific library. As revealed by prior studies,  hallucinations are common in code generation~\cite{kabir2024zs4czeroshotsynthesiscompilable,spracklen2024we}, such as generating non-existing and fake APIs. Therefore, we then examine if the recommended APIs by LLM are deemed in the class. We design our prompt template for Task 1 as follows. 

\begin{tcolorbox}[colback=black!5!white,colframe=black!75!black, before upper={\parindent15pt\noindent}, title=Prompt Template for Task 1]
{\footnotesize
 \underline{\textbf{\@\@Instruction:}} \\
I want to use \{packageName.className\} class from Java. Recommend a list of useful with at most \{snippet\_number\} API method for this class, excluding method inherent from its parent class. For each API method specify its return type and parameters in the below format: \\
``API signature'': Description of the API
\\
\\
\underline{\textbf{For example:}} \\
``boolean add(E e)'': This method appends the specified element to the end of this list.\\
...\\
\underline{\textbf{Response:}}

 }
\end{tcolorbox}


We keep the template simple to avoid fluctuations in results caused by prompting techniques on different LLMs. We provide an example in the context (i.e., one-shot) to help LLM understand our intent and desired format for output. We initially tried a zero-shot prompt, but the LLM did not consistently follow the instructions well, leading us to adopt the one-shot strategy. When prompting following this template, we replace the placeholder \{packageName.className\} with the actual full class name for which we seek a recommendation. To facilitate automated analysis and evaluation, we instruct the LLM to return API recommendations in a specific format, as shown in the template, enabling easy extraction using regular expressions. To reinforce this format, we include an example that demonstrates the desired output. We also use the placeholder \{snippet\_number\} to limit the number of APIs recommended by the LLM. In this study, we set \{snippet\_number\} to five. Although we set up a threshold, LLMs possibly generate more APIs than we ask for. sometimes LLMs  After the LLM generates its response, we use regular expressions to extract the signatures of all recommended APIs for further analysis.

\subsubsection{Task 2: API-oriented Code Example Generation}\label{sec:task2}
For Task 2, we instruct the LLM to generate a code example for a given API (i.e., method). We present the prompt template for Task 2 below:

\begin{tcolorbox}[colback=black!5!white,colframe=black!75!black, before upper={\parindent15pt\noindent}, title=Prompt Template for Task 2]
{\footnotesize
 \underline{\textbf{\@\@Instruction:}} \\
I want to learn how to use \{\textit{method}\} from \{\textit{packageName.className}\}. Generate a complete code example of this method. The code example needs to be executable with import statement and put the method and code snippet in the format below:\\
\textit{Code snippet:} \\
public class Main \{

public static void main(String[] args) \{ \\

\} \\
\} \\
\underline{\textbf{For example:}} \\
boolean add(E e): This method appends the specified element to the end of this list.\\
\textit{Code snippet:}\\
import java.util.ArrayList;\\
public class Main \{

public static void main(String[] args) \{

ArrayList<String> list = new ArrayList<>();

list.add("Hello");
        
System.out.println(list);

    \}\\
\}\\
\underline{\textbf{Response:}}

 }
\end{tcolorbox}

In this template, similar to the one used in Task 1, we prompt the LLM to generate a code example for a given API (i.e., \{method\}) in a package (i.e., \{package.className\}). We ask the LLM to output an executable code example with the necessary import statement and follow a specific format shown in the template. Like Task 1, this design ensures that we can extract the code examples from the response using regular expressions. However, our observations show that providing instructions alone is insufficient to consistently enforce the desired output format. Therefore, we provide an example of the format we hope the LLM will follow in the ``For example'' section to improve the likelihood of generating code in a format as we expected. We use regular expressions to extract code examples from the LLM's response for further analysis.



\subsection{Evaluation Metrics}\label{sec:evaluationMetrics}

For Task 1, our goal is to assess whether LLMs can accurately recommend APIs within a specified package. To evaluate the quality of these recommendations, we compute the proportion of incorrectly recommended APIs relative to the total recommended APIs, denoted as \textbf{\NoneExistP}. This metric is calculated as $\frac{|IncorrectAPIs|}{|All\ recommended\ APIs|}$. A lower value indicates that the LLM provides more accurate API recommendations, reflecting higher quality. 
A recommended API is considered correct only if it exists within the specified package and all elements of its signature (i.e., return type, method name, number of parameters, and parameter types) match exactly with the corresponding API in the package. To compute this metric, we cross-reference the actual APIs in the package to ensure that the recommended APIs not only exist but also have signatures that precisely match our records. It is important to note that if an LLM recommends incorrect APIs, this can be considered a form of hallucination, as the recommended APIs do not exist within the package. 

For Task 2, we propose three metrics to evaluate the quality of the generated code examples, each capturing a different type of error: 
\begin{itemize}[leftmargin=0.2in]
    \item \textbf{\NoAPIUsedP:} This metric assesses whether the specified API is invoked in the generated code example. When the LLM fails to include the requested API, it be categorized as instruction inconsistency hallucinations, where the model does not follow the given instruction~\cite{hallucination-categorization}. For instance, consider a scenario where the model is asked to generate a code example for the `getDecoder()' method from `java.util.Base64', but the method is absent in the generated code. This situation would be classified as a \textbf{NoAPIInvoked} case. The metric is calculated as the proportion of examples where the specified API is not invoked, defined as $\frac{|NoAPIInvoked|}{|All\ recommended\ Code|}$, where $|All\ recommended\ Code|$ represents the total number of generated examples, and $|NoAPIInvoked|$ is the number of examples where the API is missing. 
    \item \textbf{\UnCompilableP:} After excluding the \textit{NoAPIUsed} cases, we examine whether the generated code examples can be compiled successfully. We calculate the proportion of code examples that are not compilable with compilation errors, denoted as \textbf{\UnCompilableP}, which is defined as $\frac{|UnCompilable|}{|All\ recommended\ Code|}$, where $|UnCompilable|$ is the number of uncompilable code examples. 
    \item \textbf{\NonRunnableP:} We calculate the proportion of code examples that cannot be executed with runtime errors while can be compiled successfully, be denoted as \textbf{\NonRunnableP}. It is calculated as $\frac{|Unexecutable|}{|All\ recommended\ Code|}$, where $|Unexecutable|$ is the number of Unexecutable code examples. 
    
\end{itemize}

In Task 2, we do not evaluate the correctness of the generated code examples with test cases, as we prompt LLMs to only generate code examples. Instead, we rely on our metrics, i.e., \NoAPIUsedP, \UnCompilableP, and \NonRunnableP—whose sum, denoted as \textbf{\total}, serves as a \textbf{lower bound} for evaluating the quality of the generated code. This is because functionally correct API-oriented code must, at the very least, be executable and correctly invoke the specified API. 

\section{Case Study Design}\label{sec:experimentalsetting}

In this section, we illustrate a case study, where we utilize our framework \ourtool to perform an empirical study to evaluate and understand API-oriented code generation by LLMs. We present our research questions (RQs), dataset, LLMs, and approach for each RQ in the following subsections. 

\subsection{Research Questions}
We conduct our case study around four research questions:

\begin{itemize}
    \item \textbf{Quality Assessment -} \rqone
    \item \textbf{Error Analysis -} \rqtwo
    \item \textbf{Factor Analysis -} \rqthree
    \item \textbf{Error Mitigation -} \rqfour
\end{itemize}

In RQ1, we investigate the quality of API-oriented code generated by LLMs, including both the recommended APIs and the generated code examples. In RQ2, we focus on understanding the types of errors that occur in the APIs and code examples produced by LLMs. This analysis provides insights into the strengths and limitations of different LLMs in generating API-oriented code. In RQ3, we explore the factors that may be associated with the quality of API-oriented code generated by LLMs. Understanding these factors can offer insights into improving LLM performance and help develop models to predict low-quality code. Lastly, Retrieval-Augmented-Generation (RAG) has been shown to enhance code generation~\cite{chen2024code,daneshvar2024exploring,parvez2021retrieval}. Accordingly, in RQ4, we investigate whether RAG can help reduce errors and improve the quality of API-oriented code generated by LLMs.


\subsection{Datasets}
In this case study, we focus on all packages from the Java Runtime Environment 8 (JRE 8), released on July 19, 2016~\cite{java-api}. We chose JRE packages due to the widespread usage of their APIs in code repositories, such as those hosted on GitHub. Since these open-source repositories are frequently used to train LLMs, it is likely that the models have substantial knowledge of these APIs. To collect all packages and APIs in each package, we crawled the API documents of JRE 8 from its official website~\cite{java-api} and stored them in our database. Our dataset consists of 217 packages, comprising 2,397 classes. For Task 1, we generate prompts based on a predefined template for each class, and for Task 2, we construct prompts using the APIs correctly recommended in Task 1.

To determine whether a code example is compilable and executable, we compile and run the Java file as a subprocess from our main script and collect any errors. Furthermore, we have implemented a timeout mechanism to terminate subprocesses that exceed a threshold. Specifically, if the code snippet runs for over 15 seconds, it is automatically terminated and marked as a failure.

\subsection{Base LLMs}
In our study, we employed three different LLMs as the base models: ChatGPT~\cite{ChatGPT}, Magicoder~\cite{wei2023magicoder}, and DeepSeek Coder~\cite{deepseekai2024deepseekllmscalingopensource}. These models were chosen due to their representation of both commercial general-purpose LLMs and open-source LLMs specialized for code tasks, as well as their ranking as top performers in code generation tasks\footnote{https://www.vellum.ai/llm-leaderboard}.

For ChatGPT, we use gpt-3.5-turbo~\cite{ChatGPT} for state-of-the-art capabilities in both general-purpose natural language processing tasks and code generation~\cite{dong2023self,yeticstiren2023evaluating}. For MagiCoder, we use Magicoder-S-DS-6.7B~\cite{magicoder}, which is trained specifically for designed for code generation and coding-related tasks. DeepSeek Coder~\cite{deepseek-coder}, built upon Deepseek-LLM 7B, was chosen for its superior ability to solve code-related tasks. For all LLMs, we set the temperature to 0.6 by following, and all other hyper-parameters were maintained at their default values. 

We conducted our analysis on all the three LLMs for RQ1. Since the performance of MagiCoder and DeepSeek Coder are similar as shown in the results of RQ1 (ref. Table~\ref{tab:RQ1}), we conducted our experiments and analysis on MagiCoder and ChatGPT for the rest RQs (RQ2-RQ4).

\subsection{Approaches for RQs}

\subsubsection{RQ1 - Quality Assessment}
In RQ1, we perform the two unit tasks on each of the 2,397 target classes extracted from JRE 8 leveraging three selected LLMs. We record the inference output from each prompt and extract the generated code. We then evaluate the quality of generated code using the metrics outlined in Section~\ref{sec:evaluationMetrics} for both tasks. We repeated each task 10 times to reduce the bias from randomness.


\subsubsection{RQ2 - Error Analysis}

For Task 1 - API recommendation, as discussed in Section~\ref{sec:evaluationMetrics}, we classify any recommended APIs that do not exist within the given package as incorrect, which can be considered hallucinations generated by the LLM. Consequently, we adopt an established categorization scheme from prior hallucination studies~\cite{hallucination-categorization} to classify the errors in API recommendations as follows:


\begin{itemize}
    \item \textbf{Factual Fabrication:} The recommended API is entirely fabricated, meaning the API name does not exist in the specified package.
    \item \textbf{Factual Inconsistency:} Unlike factual fabrication, the recommended API name does exist; however, other elements in the signature (e.g., return type or parameters) are incorrect. 
    \item \textbf{Instruction Inconsistency:} he LLM fails to follow instructions regarding generating a list of APIs with their full signatures. This may include missing return types, missing parameters, or generating irrelevant text instead of API signatures.
    \item \textbf{Context Inconsistency:} The LLM incorrectly claims that the context provided in the prompt is wrong and fails to follow the instructions. For example, it may claim that a specified class is not part of the package and thus does not provide any API recommendations.
\end{itemize}

We randomly selected a statistically representative sample of 384 incorrect APIs (method signatures) recommended by ChatGPT, using a 95\% confidence level and a 5\% margin of error. Given the similarity in API recommendation quality between MagiCoder and DeepSeek Coder (as shown in the RQ1 results), we focused on MagiCoder's incorrect recommendations and similarly sampled 384 incorrect APIs. Two authors independently categorized the errors in these APIs, identifying deficiencies or ambiguities by cross-referencing API documentation. They then discussed their categorizations to resolve any disagreements and reached a consensus.

In Task 2, the four types of hallucinations defined in Task 1 also occur. For example, if a fabricated API is used in a generated code example, we classify it as a Factual Fabrication. Similarly, if the LLM fails to invoke the specified API, resulting in a NoAPIInvoked case, we label it as Instruction Inconsistency. However, in Task 2, a code example could still have compilation or runtime errors without hallucinating on APIs. In other words, the four types of hallucination errors cannot cover all cases. To address this, we conducted an open coding process to derive additional error categories (subcategories under compilation errors and runtime errors), following the methodology of prior studies~\cite{zhang2019empirical,wu2019developers,seaman1999qualitative}. We began by randomly sampling 384 unexecutable code examples generated by MagiCoder. The first two authors manually reviewed 100 randomly selected examples from this set to develop an initial list of error categories. During this process, the categories were iteratively refined and revised. Once the error types were finalized, both authors independently applied these categories to the remaining 284 samples. After labeling, they discussed their findings to resolve disagreements and reached a consensus. The final list of derived error types is shown in Table~\ref{tab:task2ErrorTypes}. We followed the same process for ChatGPT, i.e., randomly sampling 384 unexecutable code examples and categorizing them accordingly. It is important to note that a code example may be unexecutable or uncompilable due to hallucinations, such as invoking fabricated APIs. In such cases, we categorize it as a hallucination rather than a Runtime Error or Compilation Error. 

\begin{table}[]
\caption{Types of errors occurred in generated code examples by LLMs.}\label{tab:task2ErrorTypes}
\scriptsize
\begin{tabular}{p{0.6in}|p{0.8in}|p{3.5in}}
\hline
\textbf{Type}                                         & \textbf{Sub-Type}         & \textbf{Definition} \\ \hline

\multirow{3}{*}{\parbox{0.6in}{Hallucination}}  
    & Factual Fabrication       &   A fabricated API was invoked in the code example. 
     \\ \cline{2-3} 
    & Factual Inconsistency     &   Different from Factual Fabrication. The name of the recommended API does exist, while other elements in the signature (i.e., return type and parameters) are wrong.  \\ \cline{2-3} 
    & Instruction Inconsistency &   The LLM does not follow the instructions to generate code examples, such as \textit{NoAPIInvoked} cases.                  \\ \cline{2-3} 
    & Context Inconsistency &  The LLM indicates that the context provided in the prompt is wrong and does not follow instructions. For example, the LLM indicates that the specified API is not in the given package and does not generate any code.              \\ \hline
                             
\multirow{5}{*}{\parbox{0.6in}{Compilation Errors}}     
    & Type Mismatch             &  Errors that occurs when an operation or function receives a variable or argument of a different data type than expected (e.g., assigning an int to a string variable.).                   \\ \cline{2-3} 
    & Missing Import Statement  &  Errors are caused by missing import statements in the code example. \\ \cline{2-3} 
    & Polymorphism Error        &  Error that are related to polymorphism. (e.g., does not fully implement an abstract method from a superclass). \\ \cline{2-3} 
    & Undeclared variable/class/structure      &  Error are caused due to a variable/class/structure being used without being previously declared or defined.  \\         \cline{2-3}    
    & API misuse        &  Errors occur when APIs are misused (e.g., using non-static method in a static context).  \\ \hline

\multirow{6}{*}{\parbox{0.6in}{Runtime Errors}}        

    & Initialization Error &  Errors occur when a variable/class/structure is not initialized correctly (e.g., an object is used before it has been fully or correctly instantiated or a system environment variable is not set up properly).              \\ \cline{2-3}               
    & Exception Handling  Error      &    Errors occur when Exceptions are not caught or thrown properly.                 \\ \cline{2-3} 
    & Timeout Error                  &      Errors occur when running beyond a pre-defined time (15 seconds in this study).               \\ \cline{2-3} 
    & Connection Error               &     Errors arise when the code example is unable to establish a successful connection to a remote server, service, network, or another device. \\ \cline{2-3} 
    & Missing External Resource         &  Errors occur when the code example fails to access or retrieve data from an external resource, such as a file, database, API, or network service.                   \\ \cline{2-3} 

    & API misuse       &  Errors occur when APIs are misused (e.g., using protected/private APIs, or passing the parameters in the wrong format). 
                   \\ \cline{2-3} 
    & Deprecated Error          &   Warning occurs when deprecated APIs are used. 
              \\ \hline
     
\end{tabular}
\vspace{-0.2in}
\end{table}



\subsubsection{RQ3 - Factors Analysis}



In this RQ, we aim to investigate the factors at the API level that may be associated with the quality of code generation. Specifically, for Task 1, we seek to investigate factors that are associated with whether a recommended API is correct. For Task 2, we focus on understanding the factors that are associated with the erroneous code examples (i.e., the case where is either noAPIinvoked, unexecutable, or uncompilable).

We examine these factors from two perspectives: 1) the API itself and 2) the model used. From the API perspective, we consider two factors: the API's popularity (i.e., \textbf{API\_popularity}) and its length (i.e., \textbf{API\_length}). From the model perspective, we analyze three factors: Self-Probing (i.e., \textbf{Probing}), Perplexity (i.e., \textbf{PPL}) and self-consistency (i.e., \textbf{Consistency}). Each of these factors is discussed in more detail below. 

    \textbf{API\_popularity:} LLMs are trained on public datasets, so APIs that are widely used are more likely to have substantial representation in the training data. Hence, an LLM is more likely to generate high-quality code for these popular APIs. To measure the popularity of APIs across GitHub repositories, we employ Google BigQuery~\cite{googlebigquery} to analyze the frequency of API package imports. Developers commonly import specific packages to access corresponding APIs. By querying BigQuery’s public GitHub dataset~\cite{googlebigquery-github}, we quantify API popularity by counting the number of repositories referencing each package, offering a reliable metric for API usage across GitHub.
    
    \textbf{API\_length:} Longer APIs are intuitively more challenging for LLMs to predict accurately, as there are more components that need to match exactly with the existing API. Therefore, we take the length of the API into account by considering its fully qualified name, which includes the return type, method name, and parameter types in terms of characters. 
    
    \textbf{Probing:} As noted in a previous study~\cite{hao2024empiricalstudydevelopersshared}, developers often start by asking LLMs if they are familiar with specific APIs before proceeding with detailed tasks, allowing them to probe the LLM's capacity to support their software engineering inquiries. Following this strategy, we probe the LLM to determine its knowledge of a given library. For Task 1, we ask if the LLM recognizes a specific class and request a ``Yes'' or ``No'' response by adjusting the prompt template for Task 1. Similarly, for Task 2, we ask if the LLM knows a particular API, again expecting a ``Yes'' or ``No'' response, with the value being binary (``Yes/No'').
    
    \textbf{PPL:} Perplexity quantifies a model's uncertainty when generating text, with a lower value suggesting more accurate predictions that align closely with the actual text  distribution~\cite{chen1998evaluation}. To compute perplexity, we enabled the LLMs to return output tokens and log probabilities, where the log probability reflects the likelihood of a token appearing in a sequence given its preceding context. Perplexity is calculated by exponentiating the negative average of these log probabilities. For Task 1, we computed the perplexity for each returned API, and for Task 2, we calculated it for the returned code snippets, providing a measure of the model's confidence. 

    \textbf{Consistency:} Self-consistency measures the certainty or reliability an LLM demonstrates through its internal coherence across multiple responses to the same or similar inputs~\cite{wang2022self}. Higher self-consistency usually indicates greater confidence in generating outputs aligned with factual knowledge. For Task 1, we ran each prompt 10 times and determined self-consistency by calculating the frequency of each API's occurrence; a higher frequency indicates greater consistency (e.g., an API appearing 8 out of 10 times has a self-consistency of 0.8). For Task 2, we repeated the prompt for each API 10 times and measured consistency using the distance to center, which quantifies intra-similarity among the generated code examples~\cite{rajaraman2011mining}. To measure the distance of two code examples, we embedded examples into vectors using Sentence Transformer~\cite{sentence-transformers}. A smaller distance indicates a code example is more consistent with others. 

After collecting the factors, we divided the APIs and code examples into two groups. For Task 1, we categorized the APIs based on whether they were recommended correctly, as described in Section~\ref{sec:evaluationMetrics}: correctly recommended APIs (\textit{API$_{correct}$}) and incorrectly recommended APIs (\textit{API$_{incorrect}$}). For Task 2, we grouped the code examples into two classes: erroneous code examples (\textit{Code$_{erroneous}$}) and non-erroneous code examples (\textit{Code$_{non-erroneous}$}). To determine whether the studied factors differ significantly between the two groups in both tasks, we employed the Mann-Whitney U test, a non-parametric test that does not require any assumptions about the underlying data distribution. Additionally, we computed Cliff’s d to measure the effect size of the differences between the two groups, indicating the magnitude of the difference. The interpretation of effect size follows the thresholds provided by Cliff~\cite{cliff1993dominance}: |d| < 0.147 indicates a negligible effect, |d| < 0.33 indicates a small effect, |d| < 0.474 indicates a medium effect, and values larger than 0.474 indicate a large effect. A factor with a significant difference and non-negligible effect size between the two groups indicates this factor is a good indicator of the difference between the two groups.

We also built classification models to determine whether it is feasible to predict incorrect APIs and erroneous code examples for Tasks 1 and 2, respectively, using the proposed factors. We split the dataset into an 80:20 ratio for training and testing, following prior studies~\cite{yang2023does,ahmed2024studying,xu2017answerbot,qiao2020deep}. We selected Random Forest as the classifier for both tasks due to its generally high accuracy~\cite{rajbahadur2017impact,ghotra2015revisiting,yang2024simclone} and its robustness to outliers~\cite{cutler2012random}. To evaluate the performance of the models, we measured precision, recall, and F1-score. Additionally, we assessed feature importance~\cite{menze2009comparison} to understand the contribution of each factor to the prediction following the approach of prior studies~\cite{rajbahadur2019impact,wang2023study,santos2020predicting}. Note that to mitigate the bias from highly correlated factors, we computed the correlation among all studied factors, and observed that those factors present various ranges of correlation. For instance, model-related factors PPL, Probing, and Consistency share relatively high correlations, ranging from 0.14 to 0.59. API-related factors API\_length and API\_popularity are correlated with a range from 0.44 to 0.51. According to prior studies~\cite{rajbahadur2019impact,wang2023study,santos2020predicting}, factors with a correlation of 0.7 are considered highly correlated and should select one among them as the representative. In our case, no pairs of factors exceed this threshold, so we keep all of them. 

Similar to RQ2, we conduct our experiments on ChatGPT and MagiCoder for this RQ. 


\subsubsection{RQ4 - Error mitigation}

Retrieval-augmented generation (RAG) has been shown to enhance code generation by integrating relevant external knowledge into the language model's context~\cite{chen2024code,parvez2021retrieval,daneshvar2024exploring,lu2022reacc,tan2024prompt,nashid2023retrieval}. We aim to investigate whether employing RAG can improve API recommendation and code example generation.
For Task 1, we enrich the context provided to the LLM by including descriptions of both the package and the class in front of the ``Instruction'' section in the prompt template, as detailed in Section~\ref{sec:task1}. We denote this RAG strategy as \textbf{RAG$^{desc}_{T1}$}. For example, when requesting API recommendations for the class ``HashTable'' in the “java.util” package, we prepend the prompt with relevant descriptions: ``Package description: {package description of java.util}; Class description: {class description of HashTable}''. In addition, we explore a variant of \textbf{RAG$^{desc}_{T1}$}, where we add a list of existing APIs within the class as the additional context to test if this further improves the LLM's ability to recommend correct APIs, when the existing correct APIs are actually provided. We denote this RAG strategy as \textbf{RAG$^{desc+API}_{T1}$}. In this study, we provide a list of 10 APIs.
For Task 2, we extend the context used in Task 1 by adding a detailed description of the specific API, including its summary, return type, and input parameters. Our goal is to determine if this additional information improves the quality of code example generation. We denote the RAG strategy as \textbf{RAG$^{desc}_{T2}$}.

\section{Results of Case Study}\label{sec:results}
\subsection{RQ1 - Quality Assessment}\label{sec:rq1}


\begin{table}
\centering
\footnotesize
\caption{The quality of API-oriented code generated by the three studied LLMs, MagiCoder, DeekSeek Coder, and ChatGPT for both tasks. The cells with the best result are marked in bold.}\label{tab:RQ1}
\resizebox{\columnwidth}{!}{
\begin{tabular}{l|c|c|c|c|c}
\hline
\multirow{2}{*}{\textbf{Model}} & \textbf{Task 1} & \multicolumn{4}{c}{\textbf{Task 2}}                                                          \\ \cline{2-6} 
   & \textbf{\NoneExistP}  & \textbf{\NoAPIUsedP} & \textbf{\UnCompilableP} & \textbf{\NonRunnableP} & \textbf{\total}\\ \hline
MagiCoder              & 84.1\%           & 20.7\%              & 22.4\%           & \textbf{11.4\%}   &  54.5\%   \\
DeepSeek Coder         & 82.9\%          & 9.9\%               & 25.5\%          & 15.7\% &  51.1\%  \\
ChatGPT                & \textbf{58.1\%}           & \textbf{5.4\%}               & \textbf{20.4\%}            & 13.6\%   &  \textbf{39.4\%}   \\ \hline
\end{tabular}
}
\vspace{-0.1in}
\end{table}


\textbf{Hallucinations are prevalent in the API recommendation task, with 58.1\% to 84.1\% of the recommended APIs not existing in the specified package.} Table~\ref{tab:RQ1} summarizes the quality of recommended APIs for Task 1. Specifically, 84.1\%, 82.9\%, and 58.1\% of the recommended APIs from MagiCoder, DeepSeek Coder, and ChatGPT, respectively, do not exist in the specified package. In addition, we analyze the errors that occurred within incorrect API recommendations by analyzing where the errors occurred in the method signature. Table~\ref{tab:rq1_task1_error} presents the types of errors produced by the three LLMs. 1.8\% - 15.6\% of the errors are due to not recommending method APIs (i.e., NotMethod), where LLMs suggested other class elements, such as fields, instead of methods in these cases. The majority of the errors involved recommending method names that do not exist. The remaining errors were attributed to incorrect return types or parameters (Incorrect ReturnType/Parameter). Notably, we only assessed the correctness of return types and parameters when the method name was accurate. Therefore, it is possible that multiple errors could occur in different parts of a single recommendation.
Interestingly, among the Incorrect Return/Parameter cases, a remarkable portion of errors (85.6\% - 86.2\%) resulted from combining return types and parameters from multiple overloaded methods. For example, the method "boolean remove(Object o)" was recommended by MagiCoder for the java.util.Hashtable class. However, only two overloaded methods ``V remove(Object key)'' and ``boolean remove(Object key, Object value)'' exist in this class. This is likely because of the common scenario of method overloading in Java, making LLMs confused when recommending API methods.

\begin{table}[h]
\footnotesize
\caption{The types of errors occurred in the recommended APIs by MagiCoder, DeepSeek Coder, and ChatGPT.}\label{tab:rq1_task1_error}
\resizebox{\columnwidth}{!}{
\begin{tabular}{l|c|c|c}
\hline
Model          & NotMethod       & MethodNameNotExist     & Incorrect ReturnType/Parameter \\ \hline
MagiCoder      & 15.6\% & 77.0\% & 7.4\%           \\
DeepSeek Coder & 10.0\% & 81.5\% & 8.5\%            \\
ChatGPT        & 1.8\%     & 77.9\%   & 20.3\%         \\ \hline
\end{tabular}
}
\end{table}

\textbf{39.4\% to 54.5\% of LLM-generated code examples have errors. More specifically, a range from 5.4\% to 20.7\% of the code examples fail to include the API they are intended to demonstrate and the rest of the generated code examples fail to compile or execute.} As shown in Table~\ref{tab:RQ1}, 5.4\% to 20.7\% of the recommended code examples omit the specified APIs entirely, which is a form of hallucination (see details in RQ2). In these cases, the LLM fails to follow the instructions to generate code examples for the given API. 20.4\% to 25.5\% of the generated code examples fail to compile and 11.4\% and 15.7\% of the code examples fail to execute successfully. 


\textbf{Different LLMs vary in performance across different tasks. Compared with MagiCoder and DeepSeek Coder, ChatGPT tends to follow the instructions better and have a lower rate of hallucinations, while producing a similar proportion of uncompilable/unexecutable code examples.} 
For Task 1, ChatGPT generated a significantly lower proportion of incorrect APIs than MagiCoder and DeepSeek Coder. Furthermore, MagiCoder and DeepSeek Coder tended to produce more APIs than ChatGPT. Although the LLMs were instructed to recommend a maximum of 5 APIs per prompt (totaling 11,9850 APIs for 2,397 classes with 10 repetition), MagiCoder and DeepSeek Coder often exceeded this limit, whereas ChatGPT followed the instructions more accurately. For instance, ChatGPT generated 114,339 APIs while MagiCoder generated 153,470 for Task 1.
A similar trend was observed in Task 2, where ChatGPT had fewer NoAPIInvoked cases than MagiCoder and DeepSeek Coder.
However, MagiCoder, DeepSeek Coder, and ChatGPT share similar performance in generating compilable and executable code.




\rqboxc{Different LLMs vary in performance across different tasks. Compared with MagiCoder and DeepSeek Coder, ChatGPT tends to follow the instructions better and have a lower rate
of hallucinations, while producing a similar proportion of uncompilable/unexecutable code examples. Hallucinations are prevalent in the API recommendation task, with 581.\% to 84.1\% of the recommended APIs not existing in the specified package. For code example generation, 39.4\% to 54.4\% of code examples have errors. More specifically, a range from 5.4\% to 20.7\% of the code examples fail to include the given API, and the rest of the generated code examples fail to compile or execute. }

\subsection{RQ2 - Error Analysis}\label{sec:rq2}

\begin{table}[]
\caption{The count and distribution of each type of error in Task 1 and Task 2. }\label{tab:rq2ErrorTypes}
\footnotesize
\begin{tabular}{p{0.8in}|p{1.7in}|c|c|c|c}

\hline
\multicolumn{1}{l|}{\multirow{2}{*}{\parbox{1in}{\textbf{Type}}}}  & \multirow{2}{*}{\textbf{Sub-Type}}                                                      & \multicolumn{2}{c|}{\textbf{MagiCoder}}                & \multicolumn{2}{c}{\textbf{ChatGPT}}                  \\ \cline{3-6} 
                           &                                                                                         & \multicolumn{1}{l|}{\textbf{Task 1}} & \textbf{Task 2} & \multicolumn{1}{l|}{\textbf{Task 1}} & \textbf{Task 2} \\ \hline

\multirow{4}{*}{\parbox{1in}{Hallucination}}                       
& Factual Fabrication     & 180 (46.8\%)            & 52 (13.5\%)              & 197 (51.3\%)             & 46 (12.0\%)              \\ \cline{2-6} 

& Factual Inconsistency   & 89 (23.2\%)              & 55 (14.3\%)             & 66 (17.2\%)             & 42 (10.9\%)           \\ \cline{2-6} 
& Instruction Inconsistency    & 115 (30.0\%)            & 27  (7.0\%)            & 121 (31.5\%)            & 6 (1.6\%)              \\ \cline{2-6} 
& Context Inconsistency      &           N/A             & 6 (1.6\%)     &       N/A         &      0           \\ \hline
                                                      
\multirow{4}{*}{\parbox{1in}{Compilation Errors}}      
& Missing Import Statement       &       N/A       & 19 (4.9\%)              &   N/A  & 32  (8.3\%)            \\ \cline{2-6} 
 & Type Mismatch     &  N/A       & 21 (5.5\%)              &   N/A & 9 (2.3\%)            \\ \cline{2-6} 
 & Polymorphism Error  &     N/A     & 22 (5.7\%)      &        N/A & 16 (4.2\%)             \\ \cline{2-6} 
 & Undeclared variable/class/structure       &      N/A    & 3 (0.8\%)              &  N/A  &        0        \\ \cline{2-6} 
 & API misuse  &      N/A    & 6 (1.6\%)              &  N/A  &        52 (13.5\%)        \\ \hline

\multirow{5}{*}{\parbox{1in}{Runtime Errors}}         
& Initialization Error &  N/A              & 68 (19.3\%)             &  N/A    & 106 (27.6\%)            \\ \cline{2-6} 
 & Exception Handling Error       &  N/A  & 22 (5.7\%)   &   N/A & 20 (5.2\%)             \\ \cline{2-6} 
& Timeout Error     &  N/A & 11 (2.9\%)             &     N/A & 1 (0.3\%)              \\ \cline{2-6} 
& Connection Error   &    N/A  & 16 (4.2\%)             &  N/A  & 3 (0.8\%)              \\ \cline{2-6} 
& Missing External Resource  &   N/A    & 25 (6.5\%)     &    N/A & 20 (5.2\%)              \\ \cline{2-6} 
                                                      
& API Misuse  &   N/A & 21 (5.5\%)              &   N/A & 10 (2.6\%)             \\ \cline{2-6} 
& Deprecated Error         &  N/A & 9 (2.3\%)    &  N/A  & 23 (6.0\%)              \\ \hline
\end{tabular}
\vspace{-0.2in}
\end{table}

\textbf{For the API recommendation task, most errors are due to Factual Fabrication Hallucinations (46.0\%/51.3\%), followed by Instruction Inconsistencies (30.0\%/31.5\%), and finally, Factual Inconsistencies for both MagiCoder and ChatGPT.} Table~\ref{tab:rq2ErrorTypes} presents the distribution of error types for Task 1. A similar trend is observed for both MagiCoder and ChatGPT. The majority of errors stem from Factual Fabrication. For example, when asked to recommend APIs for ``java.util.Arrays'', the LLM suggested ``createCompatibleGraphics()'', which does not exist. Upon reviewing the API documentation, we found a similar method, ``createCompatibleImage()'', indicating that the LLM likely confused the terms `Graphics'' and ``Image'' after generating the prefix ``createCompatible''. The second most common error is Instruction Inconsistency. For instance, the LLM often disregards the required format, omitting return types or parameters for certain static APIs. The least frequent error is Factual Inconsistency, where the LLM generates the correct API name but with incorrect return types or parameters.


\textbf{In the code example generation task, the most common error categories are runtime errors (46.4\%/47.7\%), followed by hallucinations (36.1\%/24.5\%) and compilation errors.} As presented in Table~\ref{tab:rq2ErrorTypes}, Runtime Errors occur the most frequently. 46.4\% and 47.7\% of the errors occur during runtime for MagiCoder and ChatGPT, respectively. In general, those two LLMs share similar patterns, the most frequent errors are Initialization Error. For instance, a code example for ``getClickCount()'' in the package ``java.awt.event.MouseEvent'' was generated by MagiCoder as shown in Listing~\ref{rq2-contextPreparation}, null was passed as the Component source parameter for ``MouseEvent()''. The source should not be null, it should be a valid component (like a JFrame or a JButton). The code was not initialized properly. Other examples of this type of error could relate to system environment preparation/configuration (e.g., system environment variable).

\begin{lstlisting}[
    float,
    basicstyle=\footnotesize,
    language=Java,
    caption=Code example of Context Initialization Error,
    label=rq2-contextPreparation
]
import java.awt.event.MouseEvent;

public class MouseEvent_3 {
    public static void main(String[] args) {
        MouseEvent event = new MouseEvent(null, 0, 0, 0, 0, 0, 0, false, 0);
        int clickCount = event.getClickCount();
        System.out.println("Number of times the mouse button has been clicked: " + clickCount);
    }
}
\end{lstlisting}

Hallucinations accounted for 36.1\% and 24.5\% of the errors in Task 2 when using MagiCoder and ChatGPT, respectively. Most of these hallucinations were categorized as Factual Fabrication and Factual Inconsistency. Unlike in Task 1, Instruction Inconsistency was rare in Task 2. Interestingly, we observed six cases of Context Inconsistency when using MagiCoder. For instance, we requested the LLM to generate a code snippet for the method ``void update(Graphics g)'' from the ``java.awt.Canvas'' package. However, the LLM incorrectly asserted that the method ``void update(Graphics g)'' is not part of Canvas package, which contradicted the context we provided. In fact, the ``java.awt.Component'' package does contain a method with the same name, ``void update(Graphics g)''. This likely confused the LLM, resulting in the hallucination.





\rqboxc{For the API recommendation task, most errors are due to Factual Fabrication Hallucinations (46.0\%/51.3\%), followed by Instruction Inconsistencies (30.0\%/31.5\%), and finally, Factual Inconsistencies for both MagiCoder and ChatGPT. For code example generation, the most errors occur due to Runtime Errors (e.g., Context Initialization Error), followed by Hallucination and Compilation Errors.}








\subsection{RQ3 - Factors Analysis}\label{sec:rq4}


\begin{table}[]
\caption{Statistic test results for Task 1 and Task 2.}\label{tab:rq4-task1-stat-test}
\resizebox{\columnwidth}{!}{
\begin{tabular}{l|l|l|c|c|c|c}
\hline
 & \textbf{Model }                     & \textbf{Factor}          & \textbf{Mean of API$_{correct}$} & \textbf{Mean of API$_{incorrect}$} & \textbf{p-value} & \textbf{Cliff's D}        \\ \hline
\multirow{10}{*}{Task 1} & \multirow{5}{*}{MagiCoder} & API\_popularity  & 71557.7            & 23322.5       & 0.0     & 0.490 (large)       \\
                  &         & API\_length      & 58.3               & 67.2          & 0.0     & -0.252 (small)     \\
                  &         & PPL              & 1.04            & 1.04                            & 0.34    & -0.006 (negligible) \\
                  &         & Consistency      & 0.29              & 0.14                        & 0.0     & 0.432 (medium)      \\ 
                  &         & Probing          & 0.22              & 0.11                        & 9.06e-171    &  (negligible)      \\\cline{2-7} 
 & \multirow{5}{*}{ChatGPT}   & API\_popularity  & 49,959.2           & 17,223.3                     & 0.0     & 0.360 (medium)      \\
                   &        & API\_length      & 62.4               & 75.4                          & 0.0     & 0.281 (small)      \\
                   &        & PPL              & 1.10              & 1.12                          & 2.28e-47     & -0.093 (negligible)      \\
                   &        & Consistency & 0.4545              & 0.1971                          & 0.0     & 0.487 (large)     \\  
                   &        & Probing & 0.68              & 0.47                         & 0.0     & 0.213 (small)      \\ \hline
 & \textbf{Model }                     & \textbf{Factor}          & \textbf{Mean of  Code$_{non-errornous}$} & \textbf{Mean of Code$_{errornous}$} & \textbf{p-value} & \textbf{Cliff's D}        \\ \hline
\multirow{10}{*}{Task 2} & \multirow{5}{*}{MagiCoder} & API\_popularity  & 99,549.9            & 54,182.8                        & 0.0     & 0.318(small)       \\
                  &         & API\_length      & 51.3               & 57.6                             & 2.49e-178     & -0.221 (small)     \\
                  &         & PPL              & 1.11             & 1.23                            & 8.11e-99    & -0.164 (small) \\
                  &         & Consistency & 0.28              & 0.38                         & 0.0     & -0.398 (medium) \\
                  &         & Probing & 0.65              & 0.53                        & 1.556e-77     & 0.124 (negligible)\\ \cline{2-7} 
 & \multirow{5}{*}{ChatGPT}   & API\_popularity  & 93,920.6           & 67,180.0                     & 1.59e-143     & 0.197 (small)      \\
                   &        & API\_length      & 52.5               & 58.7                          & 2.32e-74     & -0.141 (negligible)      \\
                   &        & PPL              & 1.08              & 1.15                          & 9.10e-48     & -0.112 (negligible)      \\
                   &        & Consistency & 0.21              & 0.23                        & 5.63e-38    & -0.100 (negligible)       \\ 
                   &        & Probing & 0.79              & 0.47                         & 0.0     & 0.321 (small) \\\hline                   
\end{tabular}
}
\end{table}

\textbf{For both tasks,  all studied factors show a significant difference between the two groups of generated APIs/code examples.} Table~\ref{tab:rq4-task1-stat-test} presents the statistical test results on the studied factors for Task 1 and Task 2. 
For Task 1, in both LLMs, all factors exhibit significant differences between the two groups of APIs with non-negligible effect sizes with non-negligible effect size, except Probing on MagiCoder. For example, the popularity of API$_{incorrect}$ is substantially lower than that of API$_{correct}$  with a large effect size in both LLMs. This aligns with the expectation that a more popular API, which is likely to have more related usage in LLMs' training data, increases the model's likelihood of making correct recommendations. From the model's perspective, consistency serve as a strong indicator for differentiating between the two API groups.
For Task 2, all studied factors demonstrate significant differences between the groups of erroneous and non-erroneous code examples, although API\_length, PPL, and Consistency exhibit negligible effect size on ChatGPT.

In addition, we compute the correlation between the value of \NoneExistP and each studied factor for each package. For numerical factors (i.e., API\_length, API\_popularity, Consistency, and PPL), we use the median value of all APIs as the representative for the package. For the binary factor Probing, we calculate the percentage of classes where the LLM responds ``Yes'' as the measure of self-probing for the package. For ChatGPT, the correlations between the proportion of incorrect APIs and the factors are as follows: API\_length (0.35), API\_popularity (-0.43), Consistency (-0.59), PPL (0.22), and Probing (-0.49). Additionally, for API-related factors, we observe that packages with longer APIs and less popular packages are more likely to produce incorrect APIs. For Task 2, we compute the correlation between the total proportion of errors (\total) and the studied factors for ChatGPT. The correlations for Task 2 are as follows: API\_length (0.13), API\_popularity (-0.26), Consistency (-0.10), PPL (0.22), and Probing (-0.50). Compared to Task 1, the correlations for other factors are weaker, except for Probing. Due to space constraints, we do not present the plots for Task 2. We also observe similar patterns for MagiCoder.





\begin{figure}
\centering

  \includegraphics[width=\textwidth]{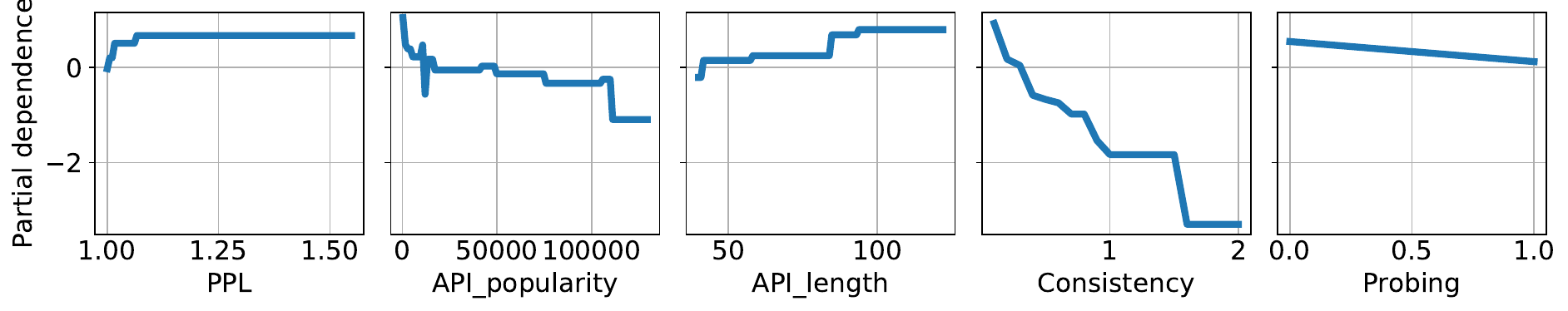}

\caption{Partial dependency plots on Task1 when using ChatGPT}
\label{fig:PDP}
\vspace{-0.1in}
\end{figure}


\textbf{Our classifiers achieve F1-scores of 0.96/0.88 for Task 1 and 0.8/0.76 for Task 2 in predicting incorrect recommended API or erroneous generated code.} Table~\ref{tab:rq4-featureImportance} presents the classification performance for both tasks. The results suggest that our proposed factors are effective indicators for distinguishing between the two groups of APIs and code examples in both tasks. We also shows the feature importance of each factor for the trained classifiers in both tasks. Factors related to Model confidence such as Consistency and PPL and API-related factors API\_propularity and API\_length serve are all important for distinguishing two groups of APIs and code examples. 

Figure~\ref{fig:PDP} demonstrates the partial dependent plots on Task1 when using ChatGPT, the result echoes our correlation analysis above. For instance, we observe the likelihood of generating an incorrect API is positively associated with PPL, API\_length, while negatively associated with API\_popularity, Consistency, and Probing. We observe similar patterns for Task 2 on ChatGPT and MagiCoder.

\begin{table}[h]
\footnotesize
\caption{The classification result and feature importance on Task 1 and Task 2.}\label{tab:rq4-featureImportance}
\begin{tabular}{c|c|ccccc|c}
\hline
\multicolumn{1}{l|}{\multirow{2}{*}{Task}} & \multirow{2}{*}{Model} & \multicolumn{5}{c|}{\textbf{Feature Importance}}                                                                                                                                                                                                                                          & \textbf{Results}           \\ \cline{3-8} 
\multicolumn{1}{l|}{}                      &                        & \multicolumn{1}{c|}{\textbf{API\_popularity}} & \multicolumn{1}{c|}{\textbf{API\_length}} & \multicolumn{1}{c|}{\textbf{PPL}} & \multicolumn{1}{c|}{\textbf{Consistency}} & \textbf{Probing} & \textbf{F1-score} \\ \hline
\multirow{2}{*}{Task1}                     & MagiCoder              & \multicolumn{1}{c|}{0.216}                                                              & \multicolumn{1}{c|}{0.297}                                                          & \multicolumn{1}{c|}{0.341}        & \multicolumn{1}{c|}{0.133}                & 0.013            & 0.96              \\
                                           & ChatGPT                & \multicolumn{1}{c|}{0.247}                                                              & \multicolumn{1}{c|}{0.221}                                                          & \multicolumn{1}{c|}{0.328}        & \multicolumn{1}{c|}{0.183}                & 0.022            & 0.88              \\ \hline
\multirow{2}{*}{Task2}                     & MagiCoder              & \multicolumn{1}{c|}{0.264}                                                              & \multicolumn{1}{c|}{0.183}                                                          & \multicolumn{1}{c|}{0.228}        & \multicolumn{1}{c|}{0.309}                & 0.016            & 0.80              \\
                                           & ChatGPT                & \multicolumn{1}{c|}{0.256}                                                              & \multicolumn{1}{c|}{0.200}                                                            & \multicolumn{1}{c|}{0.226}        & \multicolumn{1}{c|}{0.225}                & 0.093            & 0.76              \\ \hline
\end{tabular}
\end{table}

\rqboxc{Factors such as API popularity and model confidence are strongly associated with API-oriented code quality. Using our proposed factors, we built highly accurate classifiers to detect incorrect API recommendations and unexecutable/uncompiable code examples (e.g., F1 scores of 0.96 and 0.8 for Task 1 and 2 on MagiCoder).}

\subsection{RQ4 - Error Mitigation}\label{sec:rq5}

\textbf{In general, RAG improves the quality of generated Code by LLMs, while RAG's improvements differ for different LLMs.} Table~\ref{tab:rq5} compares the quality of code generated by LLMs with and without using RAG. Across both tasks, RAG improves code quality when MagiCoder and ChatGPT are used as the base LLMs. However, the magnitude of these improvements differs between the two models. Notably, RAG brings more substantial improvements for ChatGPT than for MagiCoder. For example, in Task 2, RAG reduces \total from 44.4\% to 43.2\% for MagiCoder, a modest improvement of 2.7\%. In contrast, for ChatGPT, RAG decreases \NoneExistP from 57.3\% to 30.8\%, representing a much larger improvement of 39.6\%.

\textbf{For Task 1, it is surprising that even when provided with a list of correct APIs in the context, the LLMs still fail to recommend APIs accurately.} As shown in Table~\ref{tab:rq5}, despite having the correct APIs listed along with the context, LLMs still make a significant amount of errors in their recommendations. Specifically, 40.3\% of the APIs recommended by MagiCoder and 27.9\% by ChatGPT do not exist in the specified package. This is unexpected, as the task should be straightforward - selecting from the provided list of correct APIs. One possible explanation is that LLMs sometimes disregard the given context and rely instead on their internal knowledge that is encapsulated in the model~\cite{su2024conflictbank,marjanovic2024internal}. 

\begin{table}[h]
\caption{Comparison of the quality of API-oriennted code generation by LLMs with/without RAG. The cells with better results are highlighted in bold.}\label{tab:rq5}
\resizebox{\columnwidth}{!}{
\begin{tabular}{l|l|c|c|c|c|c}
\hline
\multirow{2}{*}{\textbf{Model}} & \multirow{2}{*}{\textbf{RAG}} & \textbf{Task 1} & \multicolumn{4}{c}{\textbf{Task 2}}                                                       \\ \cline{3-7} 
             &          & \textbf{\NoneExistP}    & \textbf{\NoAPIUsedP} & \textbf{\UnCompilableP} & \textbf{\NonRunnableP} & \textbf{\total}\\ \hline
\multirow{3}{*}{\textbf{MagiCoder}}        &   w/o RAG    & 85.3\%          & 9.1\%           & \textbf{21.0\% }         & 14.3\%          & 44.4\%         \\
 &  RAG$^{desc}_{T1/T2}$     & 84.0\%   & \textbf{7.2\%}  & 23.0\%         & \textbf{13.0\%}           & \textbf{43.2\%}        \\
  & RAG$^{desc+API}_{T1}$    & \textbf{40.3\%}  & N/A  & N/A           & N/A              & N/A          \\ \hline
\multirow{3}{*}{\textbf{ChatGPT}}         &  w/o RAG    & 57.3\%             & 4.0\%          & 32.7\%        & 14.2\%           & 51.0\%           \\
 &  RAG$^{desc}_{T1/T2}$    & 54.3\%    &\textbf{1.6\%}  & \textbf{18.6\%}  & \textbf{10.5\%}    & \textbf{30.8\%}   \\    
 & RAG$^{desc+API}_{T1}$  & \textbf{27.9\%}   & N/A  & N/A           & N/A              & N/A          \\ \hline
\end{tabular}
}
\end{table}

\rqboxc{In general, RAG improves the quality of generated Code by LLMs, while RAG's improvements differ for different LLMs. For Task 1, surprisingly, even when provided with a list of correct APIs in the context, the LLMs still fail to recommend APIs accurately.}

\section{Discussion}\label{sec:dis}
\subsection{Implications of our findings}


\textbf{Our research identifies model-related indicators for predicting incorrect API-oriented code generation by LLMs.} As shown in RQ4, our proposed factors could be used to build well-performed classifiers to identify low-quality API-oriented code generated by LLMs. More specifically, model-related factors (e.g., Consistency) are strongly correlated with the quality of APIs and code examples generated by LLMs in both tasks. The importance scores from the constructed models also highlight that PPL and Consistency are critical factors. Therefore, model-related factors could serve as indicators of an LLM's capability for API-oriented code generation for a specific library. For example, developers could directly probe the LLM by asking if it knows the library or its APIs and observe the PPL of output. Actually, we test the models that are only built with model-related factors, it achieves an F1-score of 0.96 and 0.63 for Task 1 and Task 2, respectively.

\textbf{Hallucinations are prevalent in API-oriented code generation, and future research is encouraged to mitigate these issues.} As observed in RQ1 and RQ2, various hallucinations occur across both tasks. For example, RQ1 shows that 4.0\% to 9.9\% of the generated code examples by LLMs do not include the specified APIs, consistent with findings from previous studies~\cite{spracklen2024we}. Factual Fabrication and Factual Inconsistency are the most frequent types of hallucinations, where fabricated APIs are generated. Our study suggests that these hallucinations may stem from factors such as a lack of training data and confusion over overloaded methods.
Future research should explore methods to mitigate hallucinations in API-oriented code generation. Approaches from the NLP field, like RAG, fine-tuning, and self-reflection~\cite{hallucination-categorization}, could be adapted for this context. For instance, RAG appears promising, as indicated by the results in RQ5, where it reduced errors. Another direction is enhancing self-reflection with fact-checking, as RQ4 shows that self-probing can be a good indicator of poor code generation. Additionally, API documentation and runtime results could provide valuable information for quick fact-checking of LLM-generated code~\cite{kabir2024zs4czeroshotsynthesiscompilable}. Future research could develop approaches that combine self-reflection and fact-checking to reduce hallucinations (e.g., Chain-of-Verification~\cite{dhuliawala2023chain}).



\textbf{Future research is strongly encouraged to develop more effective approaches to leverage the API-related information in RAG.} As we observed in RQ5, even providing a list of APIs in the context, LLMs cannot recommend fully correct APIs, although it reduces the proportion of incorrect APIs. One possible reason is that LLMs sometimes disregard the given context (context knowledge), and only rely on their parametric knowledge, which is encapsulated in LLM's parameters, when the context knowledge and parametric knowledge conflict as prior studies reported, typically when the prompt is long~\cite{su2024conflictbank,marjanovic2024internal,shi2023trusting}. Future research is strongly encouraged to develop more effective RAG approaches to leverage external knowledge (e.g., API documentation) to mitigate errors. For instance, approaches that align with external knowledge and emphasize context prioritization, such as faithful-to-context strategies~\cite{xu2024knowledge,zhou2023context}, could be used to guide LLMs in prioritizing contextual information.

\subsection{Threats to validity}

\noindent\textbf{Internal Validity}
Prompt engineering has a significant impact on the LLM's performance~\cite{grabb2023impact}. Different prompts probably can lead to different results. However, as we discussed in Section~\ref{sec:framework}, our tasks are basic and straightforward, LLMs usually can follow the instructions specified in prompts to complete our tasks easily as the results in Section~\ref{sec:results}.
In our framework, we propose two basic tasks, i.e., API recommendation and code examples generation, to benchmark an LLM's ability of API-oriented code generation. One threat is that an LLM's ability in our designed two tasks probably does not closely align with the LLM's ability to generate code for a specific task. However, as discussed in section~\ref{sec:framework}, the goal of our framework is to evaluate LLMs on any given library with API documentation automatically. Therefore, we do not include tasks such as code generation with specific requirements which typically need test cases in our framework. We believe our framework provides a lower boundary to assess LLMs' capability for API-oriented code generation. Nevertheless, we encourage future research to include more tasks to reflect the LLMs' ability to generate code using specific libraries with specific requirements. 
Previous studies suggest that LLM settings, such as temperature and decoding strategies, can significantly affect the quality of generated content~\cite{renze2024effect,thakur2024verigen}. In this study, we use default settings for the studied LLMs for all RQs. However, our framework enables such analysis and we examined actually whether different LLM settings such as different temperatures and different decoding strategies (i.e., beam search, top K, and greedy search) have a measurable impact on the quality of generated code (due to space limit, we do not present here). In general, a lower temperature tends to produce code of similar or higher quality for both tasks and across both LLMs. Greedy Search, Beam Search, and Top-K share similar performance. Another threat is that certain APIs are version version-sensitive. We encourage future work to take this into consideration when using our framework for evaluation.

\noindent\textbf{External Validity} relates to the generalizability of our findings. Even though we conducted our empirical study on three different state-of-the-art LLMs (i.e., ChatGPT, MagiCoder, and DeepSeek Coder) and JRE 8, our findings may not generalize well to other LLMs and libraries. We propose a framework to enable automatic and systematical analysis on other LLMs and libraries and encourage future research on more LLMs and libraries. 




\section{Related work}\label{sec:related}
\phead{Code recommendation and generation with LLM.}
In recent years, there has been increasing interest in using Large Language Models (LLMs) for generating code from natural language prompts~\cite{codexglue,codex,luo2023wizardcoder,wei2023magicoder,li2023starcoder}. Lu et al. initiated this field with CodeGPT, based on GPT-2 and specifically trained on source code~\cite{codexglue}. Chen et al. advanced this by fine-tuning GPT-3 models to create CodeX, which excels at generating both natural language and code~\cite{codex}. More recent models, such as starCoder~\cite{li2023starcoder}, WizardCoder~\cite{luo2023wizardcoder}, and MagiCoder~\cite{wei2023magicoder}, further enhance code generation capabilities. 
In addition to generating code from natural language, integrating Application Programming Interfaces (APIs) is crucial. Although a few research has explored API-oriented code generation for specific libraries~\cite{zan2023private,codegen4libs}, most of these efforts primarily focused on developing LLM-based approaches to generate code that interacts with APIs.
Several studies explored API integration during code generation and revealed issues, such as license issues~\cite{logen,librarian} and hallucination~\cite{spracklen2024we}. Different from prior studies, we focus on developing automated framework to evaluate LLMs on API-oriented code generation and enable further analysis, rather than analyzing the errors. 



\phead{Benchmarking for code generation.}
To evaluate the functional correctness of generated code, the most effective method is to test its execution against predefined test cases. Several benchmarks have been developed to assess LLMs' code generation abilities~\cite{chen2021evaluating,zhuo2024bigcodebench,zhang2023repocoder,du2023classeval,yu2024codereval}. HumanEval~\cite{human-eval} is widely used, testing code correctness through execution on Python problems. BigCodeBench~\cite{zhuo2024bigcodebench} evaluates code generation across various languages and tasks, while RepoEval~\cite{zhang2023repocoder} focuses on library-level code completion using unit tests. ClassEval~\cite{du2023classeval} challenges LLMs with class-level code generation. Unlike these benchmarks, which generally assess code generation and require test cases for evaluation, our focus is specifically on API-oriented code generation. More importantly, our proposed framework is fully automated and only relies on API documentation as the input. 

\phead{Issues with Code Generation using LLMs.}
Despite advancements in LLM-based code generation, issues such as vulnerabilities~\cite{vulner2,fu2023security,majdinasab2024assessing,fan2023large,pearce2022asleep}, compile/runtime errors~\cite{dou2024s,pan2023lost}, copyright issues~\cite{latendresse2024chatgpt} and hallucinations~\cite{liu2024exploring,dawn,nlgsurvey,spracklen2024we,codehalu} persist. For example, Pearce et al. found that around 40\% of Copilot-generated programs are vulnerable, a finding echoed by Majdinasab et al., who reported 27.25\% of code suggestions with vulnerabilities even in newer Copilot versions. Hallucinations, where LLMs generate factually incorrect content, pose challenges in producing reliable code snippets, leading to issues like intent conflicts and context deviations~\cite{liu2024exploring}. Dou et al. observed that LLMs often produce shorter but convoluted code for complex tasks, based on error types and compiler feedback~\cite{dou2024s}. Our study extends this analysis to API-oriented code generation, addressing not only hallucinations but also runtime and compilation errors. 


\section{Conclusion}\label{sec:conclusion}
We propose \ourtool, a lightweight and automated framework for evaluating LLMs in API-oriented code generation. Compatible with any library that provides API documentation, our framework focuses on two unit tasks: API recommendation and code example generation, along with four evaluation metrics, including the proportion of incorrect API recommendations and the proportion of code examples where no specific API is invoked and uncompilable/unexecutable code examples.

To demonstrate the framework's effectiveness, we conducted a case study with three LLMs ChatGPT, MagiCoder, and DeepSeek Coder on JRE 8. Our findings show notable variability in LLM performance across tasks, with ChatGPT generally following instructions better but generating more unexecutable code compared to the other models. We identify crucial factors that are associated with code quality, such as API popularity and model confidence. We develop classifiers that achieve high accuracy in detecting low-quality API recommendations and code examples. Additionally, while retrieval-augmented generation improves code quality, its effectiveness varies between different LLMs. Our findings offer valuable insights for future research directions.

\section{Data Availability}
We have made our replication package available, which contains all the code and datasets available
here~\cite{AutoAPIEval}.

\newpage
\bibliographystyle{ACM-Reference-Format}
\bibliography{main}

\end{document}